\documentclass[a4paper,11pt]{article}
\usepackage{jheppub} 
\usepackage[utf8]{inputenc}
\usepackage{subfig}

\title{%
  Generalised Known Kinematics (GKK)\\ An Approach for Kinematic Observables in Pair Production Events with Decays Involving Invisible Particles
  }
\author[a,1]{Thomas~M.~G.~Kraetzschmar,\note{Corresponding author.}}  
\author[a]{Fabian~M.~Krinner}
\author[a]{Marvin~Pfaff}
\author[b]{Navid~K.~Rad}
\author[b]{Armine~Rostomyan}
\author[c]{Lorenz~Schlechter}
\author[a,d]{Frank~Simon}
\affiliation[a]{Max-Planck-Institut für Physik, Föringer Ring 6, 80805 München, Germany}
\affiliation[b]{DESY, Notkestraße 85, 22607 Hamburg, Germany}
\affiliation[c]{Institute for Theoretical Physics, Utrecht University
Princetonplein 5, 3584 CC Utrecht, The Netherlands}
\affiliation[d]{Karlsruhe Institute of Technology, Institute for Data Processing and Electronics, Hermann-von-Helmholtz-Platz 1,
76344 Eggenstein-Leopoldshafen, Germany}

\emailAdd{kraetzsc@mpp.mpg.de}
\emailAdd{fkrinner@mpp.mpg.de}
\emailAdd{marvin.pfaff@tum.de}
\emailAdd{navid.khandan.rad@desy.de}
\emailAdd{armine.rostomyan@desy.de}
\emailAdd{l.k.schlechter@uu.nl}
\emailAdd{fsimon@mpp.mpg.de}

\input{belle2symbols.tex}
\abstract{
Missing kinematic information of known invisible particles, such as neutrinos, limit several high-energy physics analysis.
The undetected particle carries away momentum and energy information, preventing the total reconstruction of such an event.
This paper presents a new method to handle this missing information, referred to as the Generalised Known Kinematics (GKK) approach. Its event-by-event probability density distributions that describe the physically allowed kinematics of an event. For GKK, we consider the available kinematic information and constraints given by the assumed final state. 
Summing these event-wise distributions over large data sets allows the determination of parameters that influence the event kinematics. Examples are particle masses obscured by the missing information on the invisible final-state particles. The method is demonstrated in simulation studies with $\tautau$ events in $\epem$ collisions at the $\Upsilon$(4S) resonance,  presenting a new, promising approach for measuring the $\tau$ lepton mass.
}
\begin{document}
\maketitle


\section{Introduction}

\epem-colliders provide an ideal environment for precision measurements because of the well-known initial state and simpler final-state kinematics compared to hadron colliders. However, there is a class of events with non-reconstructable final state kinematics.
The existence of invisible particles causes this, most commonly neutrinos $\nu$, that escape detection, which leads to a degradation of measurement precision.
In this work, we discuss a new method to mitigate the problem of invisible particles in particle-pair events. 
We present the method using the example of $\tau$-pair production in \epem annihilation data, $\epem\to \taup \taum$. However, the method generalises to any similar problem statement.

We consider the case with boosted $\tau$ leptons that separate their decay products well in two opposite hemispheres.
For $\tau$ leptons, a decay is of the general form
\begin{equation}
    \tau \to n V + m I,
\end{equation}
with a number of $n$ visible daughters $V$ and $m$ invisible daughters $I$. There is always at least one invisible particle $I$ involved.
The existence of invisible particles $mI$ prevents us from a straightforward determination of observables in the final state, which relies on the complete kinematic information of the $\tau$. 
An example observable is the $\tau$ mass, which would be accessible via the invariant mass of the final state.

There are several approaches to tackling this problem. 
For example, ARGUS proposed an approximate method to determine the $\tau$-mass from its $\tau \to 3\pi \nu_{\tau}$ decay mode~\cite{ARGUS:1992chv}.
Using energy and momentum conservation, ARGUS derived a $\tau$ pseudo mass by neglecting the neutrino mass and approximating the flight direction of the $\tau$ as the flight direction of the  $3\pi$ (hadronic) system. Then, the pseudo mass distribution exhibits a sharp threshold behaviour in the region close to the $\tau$ mass.  
ARGUS measures the $\tau$-mass by determining the endpoint of the $\tau$-pseudo mass distribution. This method generalises to other $\tau$-decays that fulfil $n\geq 3$ and $m=1$.

The CLEO collaboration developed another possible solution~\cite{CLEO:1993vmo}, similar to considerations for the $W^{\pm}$~\cite{PhysRev.140.B721}.
The CLEO approach considers a particular class of events with one invisible particle $I$.
Kinematic considerations show that the three-momentum of the $\tau$ lies on a cone around the three-momentum vector of the respective visible system $V$.
CLEO reconstructs the $\tau$-pair kinematics from constraints given by the decay cones on both sides around the three-momentum of the visible system, assuming a mass-less $I$.
They use the simplifying assumption that the plane defined by the three-momentum vectors of $V$ and $I$ is equal for both decays of the $\tau$-pair.

Similar to \cite{KUHN1992381}, the proposed method avoids this simplification and is valid for $\tau$-pair events with at least one $\tau$ lepton decaying with one invisible particle $I$.  
We define this event to be of type $\tau_h$ (a hadronic $\tau$-decay)
\begin{equation}
    \tau_h \to n_h V_h + I_h.
\end{equation}
with $n_h \geq 1$, $V_h$ being hadrons, and $I_h$ the tau-neutrino $\nut$.
Here, we derive the probability distribution of the kinematics of $I$ and calculate the observable of interest, with a set of candidates drawn from the kinematic probability distribution of $I$ rather than a single value. 
That means we get a distribution of values instead of obtaining a single value per event. The sum of all distributions for all events accumulates to a new distribution, a limiting distribution. The limiting distribution allows determining the
observable of interest without the bias of the missing particles.
We call this approach Generalised Known Kinematics (GKK).
\section{Concept}
\label{sec:concept}
For measuring any quantity, the optimal observable is a uniform minimum-variance unbiased estimator of a statistic. The idea of GKK, as developed in this article, is inspired by the concept of such an estimator. 
For this concept to work, a statistic $T$ used to determine the estimator of interest must contain all available information~\cite{mood1974introduction}. However, particle pair events with invisible particles lose kinematic information. Due to physical constraints, we can recover some of the lost information and determine a statistic $T^{\prime}$, which is complete concerning the kinematic information.   

Let us consider a sample of $\tau$-pair events, with two hemispheres $\tau_1$ and $\tau_2$, of which we want to measure an observable dependent on the momentum-spectrum of the visible daughters in the rest-frame of $\tau_2$. 

We can identify particle-pair events by reconstructing one of the two particles undergoing a well-known decay. In this case, we demand the $\tau_1$ hemisphere be a type $\tau_h$ decay, a hadronic $\tau$-decay, which we will call tag-side.
Reconstruction of the tag-side allows studying the properties of the second particle $\tau_2$, the so-called signal-side, without introducing a reconstruction bias.

\begin{figure}[htbp]
	\centering
	\includegraphics[width=0.95\textwidth]{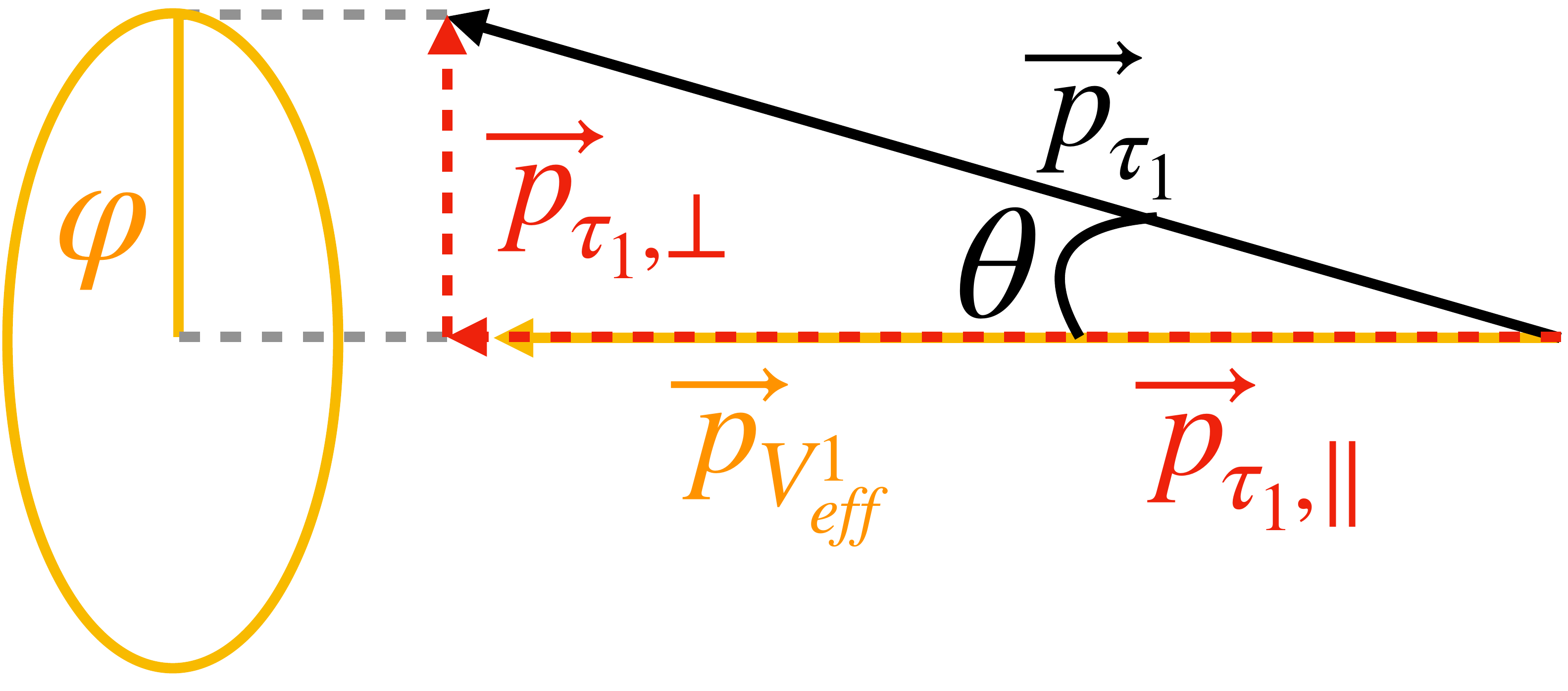}
	\caption{Visualisation of the GKK method.
             The true $\tau_1$ momentum-vector, $\vec{p}_{\tau_1}$, is shown in black. $\theta$ indicates the angle between the  $\vec{p}_{\tau_1}$ and  $\vec{p}_{V^1_{\text{eff}}}$, the resulting momentum vector sum for all visible daughters, $V^1$, is shown in yellow. The invisible particle, $I_1$, involved in the decay leads to an angular offset, which we parameterise by $\theta$ and $\varphi$. The red dashed lines indicate the components of the $\tau$-momentum. $\vec{p}_{\tau,\parallel}$ is parallel to $\vec{p}_{V^1_{\text{eff}}}$, $\vec{p}_{\tau,\perp}$ is orthogonal. The unknown component is then translated into the direction of $\vec{p}_{\tau,\perp}$, parameterised by the azimuth angle, $\varphi$.}
	\label{fig:decayStructureCirc}
\end{figure}

We simplify the $\tau_1$ decay into a two-body decay, with the invisible particle $I_1$ and the visible-daughters-system $V^1_{\text{eff}}$ by combining all $n_1 V_1$ daughters into an effective particle $V_\text{eff}^1 = \bigoplus_{i=1}^{n_1} V_1^{i}$, with its four momentum $p^{\mu}_{V_\text{eff}^1}$ given by the set of four-momentum vectors $[p^{\mu}_{V_1^1}, ..., p^{\mu}_{V_1^{n_1}}]$:
\begin{equation}
    \label{eq:obsSys}
    p^{\mu}_{V_\text{eff}^1} = \sum_{i=1}^{n_1} p^{\mu}_{V_1^i} \,.
\end{equation}
The missing information of $I_1$ translates into a probability distribution function for the $\tau_1$ momentum vector using energy-momentum conservation and the isotropic distribution of the decay of $I_1$ in the rest-frame of $\tau_1$. This isotropic distribution results in a cone-shaped momentum distribution for $I_1$ around the $\tau_1$ momentum in the centre-of-mass system of the event. Now, we can turn the argument around and constrain the $\tau_1$ 
momentum direction on a cone around the $V^1_{\text{eff}}$ direction, as shown in Figure~\ref{fig:decayStructureCirc}. 

As a first step, the $\tau_1$-energy $E_{\tau_1}$ is determined. 
In \tautau-events, neglecting initial state radiation, we approximate $E_{\tau_1}$ as half of the centre of mass beam-energy $\sqrt{s}$, 
\begin{equation}
    \label{eq:tauEnergy}
    E_{\tau_1} \approx \frac{\sqrt{s}}{2}.
\end{equation}
This approximation also allows determining the magnitude of the $\tau_1$-momentum in the centre-of-mass system via the known $\tau$-mass, $m_{\tau}$. 
Doing so, we can derive the angle $\theta$ between the $V^1_{\text{eff}}$-momentum $\vec{p}_{V^1_{\text{eff}}}$ and the $\tau_1$-momentum $\vec{p}_{\tau_1}$ by using the law of cosines\footnote{$\theta$ is given by $\cos(\theta)$ because in the polar coordinates, it is confined between 0 and $\pi$.},
\begin{equation}
    \cos(\theta) =  \frac{\vec{p}_{\tau_1} \cdot \vec{p}_{V^1_{\text{eff}}}}{|\vec{p}_{\tau_1}||\vec{p}_{V^1_{\text{eff}}}|}.
\end{equation}

The three-momentum of $\tau_1$ can be deconstructed into a parallel ($\vec{p}_{\tau_1, ||} $) and perpendicular ($\vec{p}_{\tau_1,\perp} $) component with respect to the $V^1_{\text{eff}}$ momentum direction (see Figure~\ref{fig:decayStructureCirc}). 
Constraining $\tau_1$ on a cone around $V^1_{\text{eff}}$ allows parameterising $\vec{p}_{\tau_1}$ such that the unknown direction is expressed in terms of the azimuth angle $\varphi$ in a cylindrical coordinate system parallel to $V^1_{\text{eff}}$. 

We know that each possible $\vec{p}_{\tau_1}$ on the cone is equally probable, so sampling the $\vec{p}_{\tau_1}$-distribution is simply stepping through the equally distributed $\varphi$, providing a set of equally probable candidates for the $\tau_{1}$ momentum. In contrast to other cone-based approaches such as~\cite{PhysRev.140.B721, CLEO:1993vmo}, this sampling approach is the concept of GKK.

The finite detector resolution can cause this approach to yield $\vec{p}_{\tau_1}$-momentum candidates which deviate considerably from the actual $\vec{p}_{\tau_1}$ of the event. 
This deviation makes it worthwhile to add further physical constraints. In the case of particle pair events, we can utilise the signal-side for these constraints. 

\begin{figure}[htbp]
	\centering
	\includegraphics[width=0.95\textwidth]{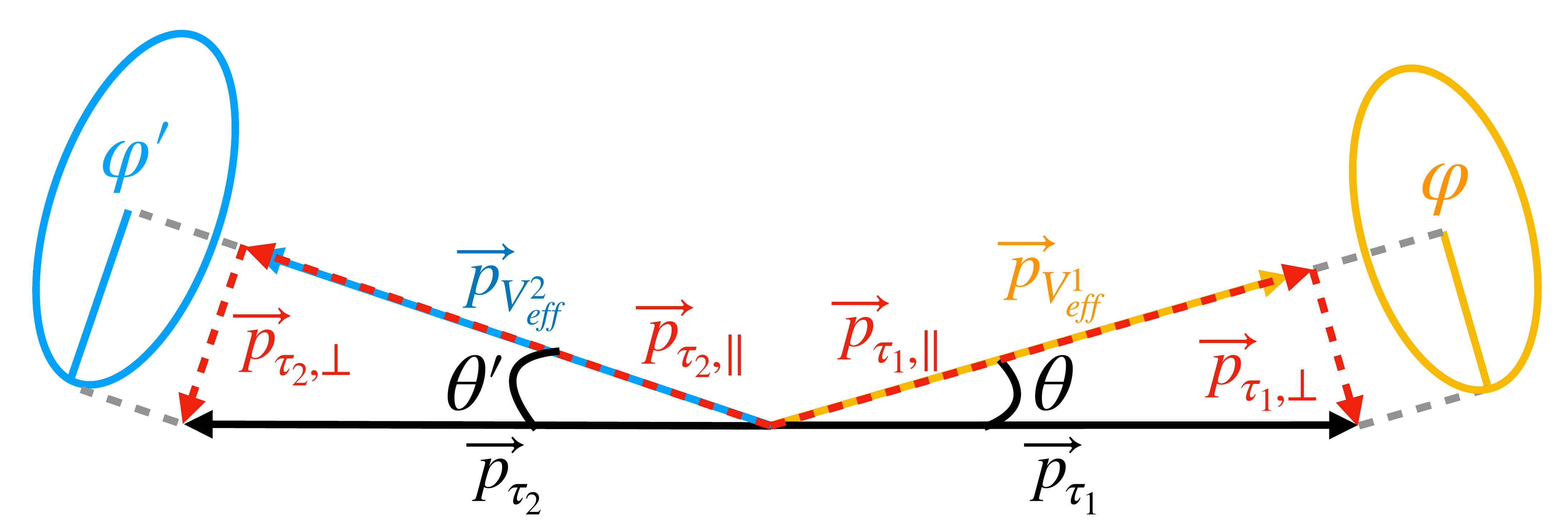}
	\caption{Visualisation of the GKK method in a special case where both taus are of type $\tau_h$. 
	         Here, we depict a \tautau event in the Centre of Mass System (CMS). 
             The true $\tau$-momentum-vectors are black. We divide the tau-pair events into a $\tau_1$- and a $\tau_2$-side. $\theta$ 
             and $\theta^{\prime}$ indicate the angle between the true $\tau$ 
             momentum vector and the sum of all reconstructed daughters on the 
             tag and signal-side, depicted in yellow and blue, respectively. 
             The $\tau_i$-momentum can be deconstructed 
             into two components indicated in red. A component that is parallel to the 
             daughter momentum $\vec{p}_{V^i_{eff},\parallel}$, $\vec{p}_{\tau_i,\parallel}$, and one 
             that is perpendicular, $\vec{p}_{\tau_i,\perp}$. 
             The unknown direction of $\vec{p}_{\tau_i}$ is translated into the direction of $\vec{p}_{\tau_i,\perp}$. The direction of $\vec{p}_{\tau_i,\perp}$ can be parametrized by the azimuth 
             angles $\varphi$ and $\varphi^{\prime}$, for $\tau_1$ and $\tau_2$ respectively.}
	\label{fig:decayStructureDoubleCirc}
\end{figure}

First, we consider a particular case and generalise afterwards. In the case of $\epem\to \tautau$ events, we can have events where both $\tau_1$ and $\tau_2$ decay hadronically. In this case, we can reconstruct the $\tau_2$ momentum similarly to $\tau_1$ with the corresponding angles $\theta^{\prime}$ and $\varphi^{\prime}$. This results in two momentum cones, as depicted in Figure~\ref{fig:decayStructureDoubleCirc}.
If the momenta of $V^1_{\text{eff}}$ and $V^2_{eff}$ were perfectly known, we could reconstruct the $\tau$-momentum by inverting the momentum on one of the two sides and looking for the momentum-vectors that fulfil the requirements imposed by energy and momentum conservation. Both $\tau$ momenta must be back-to-back in the centre-of-mass frame and lie on their respective cones. 
In general, this leads to two solutions. In extreme cases, we obtain either one solution (the cones touch each other) or infinite (the cones are on top of each other).

\begin{figure}[htbp]
    \centering
    \includegraphics[width=0.95\textwidth]{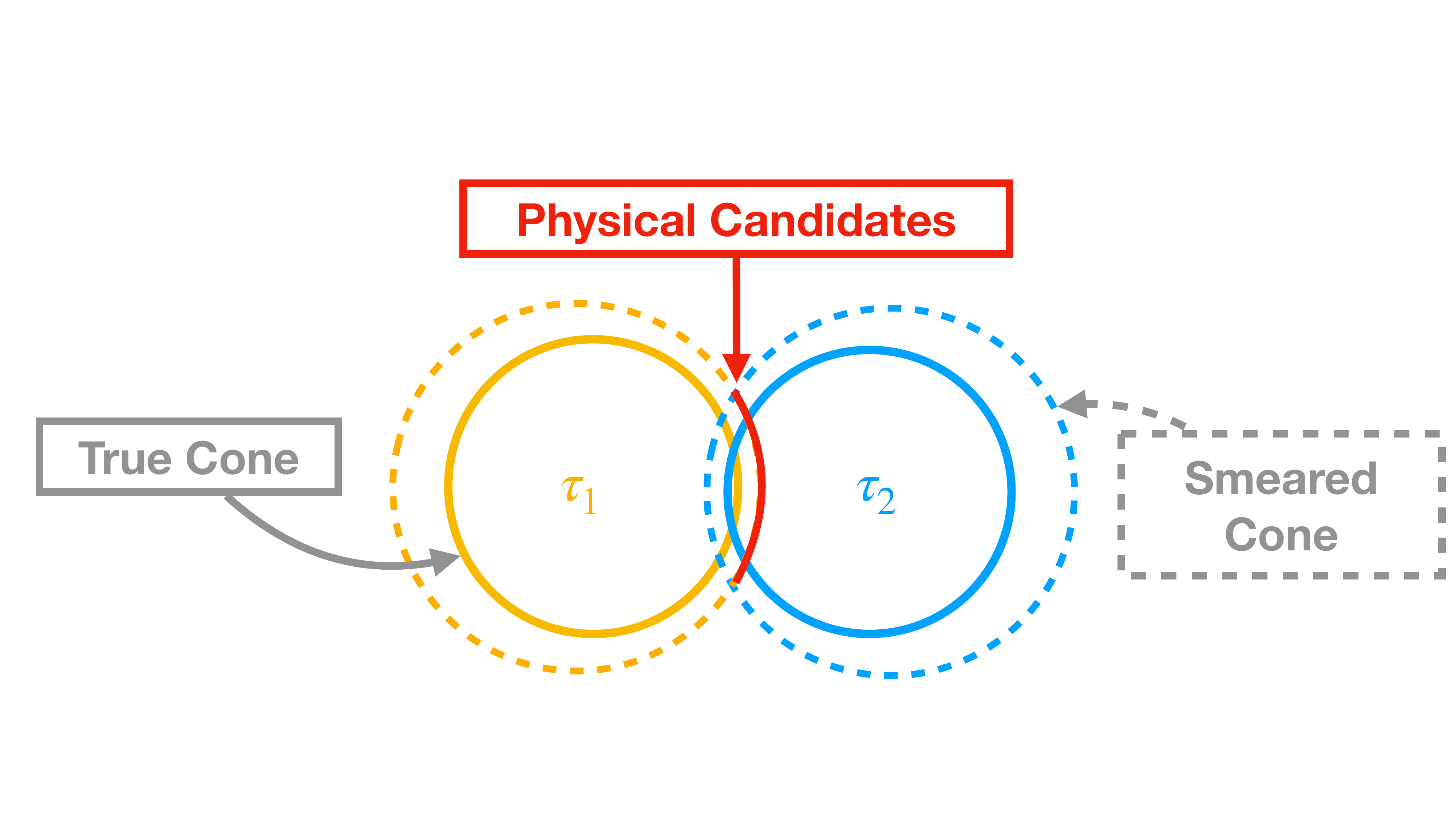}
    \caption{Sketch of detector smearing effects on the reconstructed $\tau$-momentum-candidates distributed on a cone. This example considers a $\tautau$ event where both taus are of type $\tau_h$. In an ideal case, indicated by the solid-lined circles denoted as ``True Cone'', we expect the $\tau$-cones to touch each other just at one point. The idealised case enables the determination of the true $\tau$-momentum from the common point of the $\tau$-cones. However, the cones can be wider and misaligned due to detector smearing. This case is indicated as ``Smeared Cone'' with dashed-lined circles. The smeared cones do not touch but overlap, giving rise to a physically sound range, indicated in red.}
    \label{fig:decayConeOverlap}
\end{figure}

Finite detector resolution smears the reconstructed values. This resolution effect implies that we do not have a perfect knowledge of $\vec{p}_{V^i_{eff}}$, with $i=1,2$. This deficiency causes the cones to align imperfectly but to overlap, as indicated in Figure~\ref{fig:decayConeOverlap}. 
Information loss due to $I_i$ and detector smearing cause the overlap, leading to overestimated angles $\theta$ and $\theta^{\prime}$ and slightly misaligned cones.
Instead of looking for a single $\vec{p}_{\tau}$-candidate which is in both statistics of $\tau_1$ and $\tau_2$, we are thus looking for those $\vec{p}_{\tau_1}$-candidates which are on or within the $\vec{p}_{\tau_2}$-cone, as indicated in Figure \ref{fig:decayConeOverlap}.

\begin{figure}
    \centering
    \includegraphics[width=0.95\textwidth]{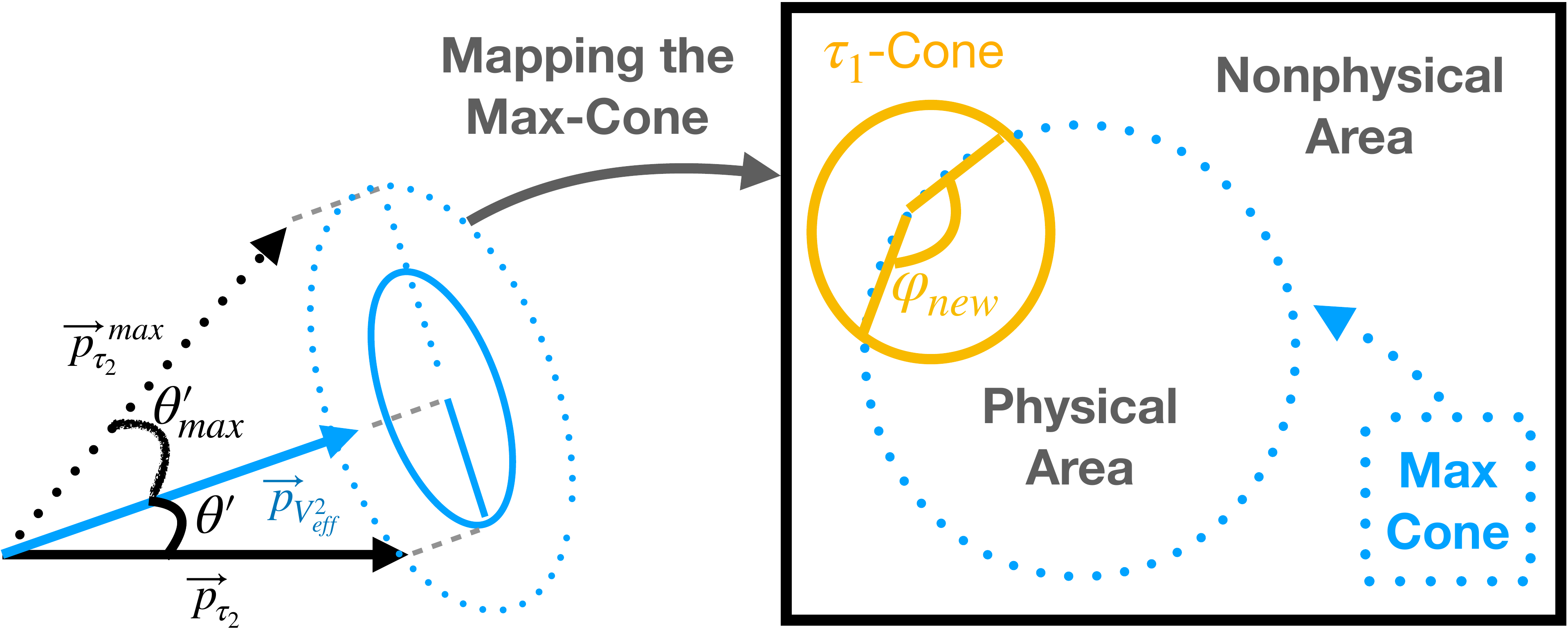}
    \caption{On the left, the construction of the maximised cone, using $\theta_{max}$ is displayed. We do this in the case of two or more invisible particles in the $\tau_2$-decay because the missing information prevents us from constructing the $\tau_2$-candidate cone. Instead, we can construct a cone with the maximum possible deviation of the $\tau_2$-momentum and the momentum of all visible $\tau_2$-daughters. We can map this cone on a 2D-plain and the cone of the $\tau_1$, which by definition has to be of type $\tau_h$. As defined in the text, the ``Max-Cone'' gives us the physical area constraining the $\tau_1$-cone to a certain range $\varphi_{new}$. We can sample this physical $\varphi$-range again, resulting in a narrower probability distribution of the $\vec{p}_{\tau_1}$.}
    \label{fig:GKKconstraint}
\end{figure}

This approach can be generalised to the case in which only $\tau_1$ decays are of type $\tau_h$, 
and $\tau_2$ has an unspecified number of invisible particles $m_2 I_2$. In this case, we cannot determine a cone of $\tau_2$-momentum-candidates. Instead, we can give a constraint to the momentum-candidate-cone of $\tau_1$ by maximising the $\theta^{\prime}$ to a $\theta^{\prime}_\text{max}$ and defining a maximised $\tau_2$-momentum-cone, as shown on the right side of Figure~\ref{fig:GKKconstraint}. How $\theta^{\prime}$ is maximised is discussed in Section~\ref{sec:math}. We constrain the $\vec{p}_{\tau_1}$-cone by demanding that it has to be within or on the cone of $\vec{p}_{\tau_2}$-candidates given by $\theta^{\prime}_\text{max}$, indicated on the left of Figure~\ref{fig:GKKconstraint}. This method rejects all $\vec{p}_{\tau_1}$-candidates outside the momentum constraints of the event. From now on, events that pass the momentum constraints are referred to as physical candidates, whereas rejected candidates are non-physical. 
As a further refinement step, we can redo the $\tau_1$ cone sampling with a restricted range of $\varphi$ to the range given by the physical candidates $\varphi_{new}$. This way, we only give weight to $\tau_1$ candidates of the physical $\varphi$-range.
%
\section{Mathematical Description}
\label{sec:math}
After we presented the concept of the GKK method in Section \ref{sec:concept}, we are now going to discuss the implementation in more detail.
We condensed the $\tau_1$-decay's visible and invisible decay products into an effective two-particle-decay problem if $\tau_1$ is of type $\tau_h$. Here, we express the invisible daughter's $I_1$ four-momentum $p^{\mu}_{I_1}$ as
\begin{equation}
    \label{eq:invis}
    p^{\mu}_{I_1} = p^{\mu}_{\tau_1} - p^{\mu}_{V_\text{eff}^1}.
\end{equation}
We can determine the angle $\theta$, displayed in Figure \ref{fig:decayStructureCirc}, by calculating the mass of $I_1$, $m_{I_1}$, with two scalar products $p^{\mu}_{I_1}\,p_{\mu, I_1}$ and  $p^{\mu}_{\tau_1} \, p_{\mu V_\text{eff}^1}$. 

First, we derive 
\begin{equation}
    \label{eq:invMass2}
    m_{I_1}^2  = m_{\tau_1}^2 + m_{V^1_\text{eff}}^2 - 2 (p^{\mu}_{\tau_1} \cdot p^{\mu}_{V_\text{eff}^1}) 
\end{equation}
to replace \textcolor{red}{} $p^{\mu}_{\tau_1} \cdot p^{\mu}_{V_\text{eff}^1}$ in 
\begin{equation}
    \label{eq:4vecDotProd}
    p^{\mu}_{\tau_1} \cdot p^{\mu}_{V_\text{eff}^1} = E_{\tau_1}E_{V^1_\text{eff}} - \cos(\theta)|\vec{p}_{\tau_1}||\vec{p}_{V^1_\text{eff}}|,
\end{equation}
which we solve for $\cos(\theta)$, resulting in
\begin{equation}
    \label{eq:cosTheta}
    \cos(\theta) = \frac{2 \, E_{\tau_1}E_{V^1_\text{eff}} + m_{I_1}^2 - (m_{\tau_1}^2 + m_{V^1_\text{eff}}^2)}{2~|\vec{p}_{\tau_1}||\vec{p}_{V^1_\text{eff}}|}.
\end{equation}
We can now study how to get $\theta_\text{max}$. To maximise $\theta$, we have to minimise $\cos(\theta)$. All components of Equation \eqref{eq:cosTheta} are given by the detected event, except for $m_{I_1}$, so we minimise $\cos(\theta)$ by setting $m_{I_1} = 0$.

The expressions derived from hereon are dependent on the reference frame; thus, the following considerations are only valid in the centre-of-mass frame. We consider all appearing quantities to be in the centre-of-mass frame. The $\tau_1$ momentum vector $\vec{p}_{\tau_1}$ is decomposed as
\begin{equation}
    \label{eq:TagMomVecDecomp}
    \vec{p}_{\tau_1} = \vec{p}_{\tau_1, \parallel } + \vec{p}_{\tau_1, \perp}\;.
\end{equation}
With the parallel component of $\vec{p}_{\tau_1}$ to $\vec{p}_{V_{\text{eff}}^1}$
\begin{equation}
    \vec{p}_{\tau_1, \parallel }= |\vec{p}_{\tau_1}|  \cos(\theta) \, \vec{n}^{\,V^1_\text{eff}}_{\parallel},
\end{equation}
and the orthogonal component of $\vec{p}_{\tau_1}$ to $\vec{p}_{V_{\text{eff}}^1}$
\begin{equation}
    \vec{p}_{\tau_1, \perp} = |\vec{p}_{\tau_1}|  \sin(\theta)\, \vec{n}^{\,V^1_\text{eff}}_{\perp},
\end{equation}
we get 
\begin{equation}
    \label{eq:TagMomVecDecompExtended}
    \vec{p}_{\tau_1} = |\vec{p}_{\tau_1}|  \cos(\theta) \, \vec{n}^{\,V^1_{\text{eff}}}_{\parallel}  + |\vec{p}_{\tau_1}| \sin(\theta) \, \vec{n}^{\,V^1_{\text{eff}}}_{\perp}.
\end{equation}
The angle $\theta$ is given by Equation~\eqref{eq:cosTheta}, $\vec{n}^{\,V^1_{\text{eff}}}_{\parallel}$ is a unit vector in the direction of $\vec{p}_{V^1_{\text{eff}}}$, and $\vec{n}^{\,V^1_{\text{eff}}}_{\perp}$ perpendicular to $\vec{p}_{V^1_{\text{eff}}}$.
We estimate the absolute value of the $\tau_1$ momentum $|\vec{p}_{\tau_1}|$ as
\begin{equation}
    \label{eq:tagMomMagEst}
    |\vec{p}_{\tau_1}| = \sqrt{\big(E_{\tau_1} \big)^2 - m_{\tau_1}^2 },
\end{equation}
using Approximation~\eqref{eq:tauEnergy} and the decay-topology.
This leaves only one unknown, the unit vector $\vec{n}^{\,V^1_{\text{eff}}}_{\perp}$. 
We can define a basis for $\vec{p}_{V^1_{\text{eff}}}$, with a parallel basis vector
\begin{equation}
    \label{eq:tagBasisPara}
    \vec{e}^{~V^1_\text{eff}}_{\parallel} = 
\begin{pmatrix}
0\\
0\\
1
\end{pmatrix}
\end{equation}
and the orthogonal basis vector
\begin{equation}
    \label{eq:tagBasisPerp}
    \vec{e}^{~V_\text{eff}^1}_{\perp} = 
    \begin{pmatrix}
        \cos(\varphi)\\
        \sin(\varphi)\\
        0
    \end{pmatrix}.
\end{equation}
In this basis, $\vec{p}_{V^1_{\text{eff}}}$ is given in z-direction. 
So, by determining the basis transformation from the detector basis to $\vec{e}^{\,V^1_\text{eff}}_{\parallel}$, 
\begin{equation}
    \label{eq:basisTrafo1}
    D_y(\rho) \cdot D_z(\xi) \cdot \vec{n}^{\,V^1_{\text{eff}}}_{\parallel} = \vec{e}_{\parallel}^{\,V^1_{\text{eff}}} ,
\end{equation}
we get an expression of $\vec{n}^{\,V^1_\text{eff}}_{\perp}$,
\begin{equation}
    \label{eq:basisTrafo2}
    \vec{n}^{\,V^1_{\text{eff}}}_{\perp} = D_z(\rho)^{\mathrm{T}} \cdot D_y(\xi)^{\mathrm{T}} \cdot \vec{e}_{\perp}^{\,V^1_{\text{eff}}}.
\end{equation}

Here, $ D_y(\rho)$ and $D_z(\xi )$ are the rotation matrices around the $y$- and $z$-axis of the detector respectively. Their definitions are
\begin{equation}
    \label{eq:Dy}
    D_y(\rho) = 
    \begin{pmatrix}
        \cos(\rho) & 0 & -\sin(\rho)\\
        0 & 1 & 0 \\
        \sin(\rho) & 0 & \cos(\rho)
    \end{pmatrix}
\end{equation}
and
\begin{equation}
    \label{eq:Dz}
    D_z(\xi) = 
    \begin{pmatrix}
        \cos(\xi) & \sin(\xi) & 0\\
        -\sin(\xi) & \cos(\xi) & 0\\
        0 & 0 & 1
    \end{pmatrix}.
\end{equation}
The angles $\rho$ and $\xi$ are the two polar angles of the laboratory frame of reference, which rotate $\vec{n}^{V^1_{\text{eff}}}_{\parallel}$ into the basis $\vec{e}^{~V^1_\text{eff}}_{\parallel}$.
By combing the above results, we obtain the expression
\begin{equation}
    \label{eq:nTagPerp}
    \vec{n}_{\perp}^{\,V^1_{\text{eff}}} = 
    \begin{pmatrix}
        \cos(\xi)  \cos(\rho) \cos(\varphi) - \sin(\xi) \sin(\varphi)\\
        \sin(\xi) \cos(\rho) \cos(\varphi) - \cos(\xi) \sin(\varphi)\\
        -\sin(\rho) \cos(\varphi)
    \end{pmatrix}.
\end{equation}
With the general expression of $\vec{p}_{\tau_1}$ and the knowledge of the distribution function $f(\varphi) = \text{const.}$, we can sample $\vec{p}_{\tau_1}$.
We obtain a statistic of size $N$ for each event, being $[^{1}p^{\mu}_{\tau_1}, ..., ^{n}p^{\mu}_{\tau_1}]$. By sampling the tag momentum with $f(\varphi)$, we obtain a statistic independent of $\varphi$, so the statistic is only dependent on the momentum of $V_{\text{eff}}^1$. 

We can use the $\tau_1$-momentum statistic to determine the $\tau_2$-daughters' momentum statistics in the $\tau_2$ rest-frame, as discussed before. The resulting distribution function of the rest-frame momentum $p^{\star}(\varphi)$ cannot be analytically inverted, as a closed-form is not known to the authors. In principle, we expect an analytical description of the limiting distribution, which could be part of future studies of the GKK method.
For the rest of this work, we consider purely numeric approaches.

\section{Results}
\label{sec:examples}

\begin{figure}[htbp]
    \centering
    \includegraphics[width=0.5\textwidth]{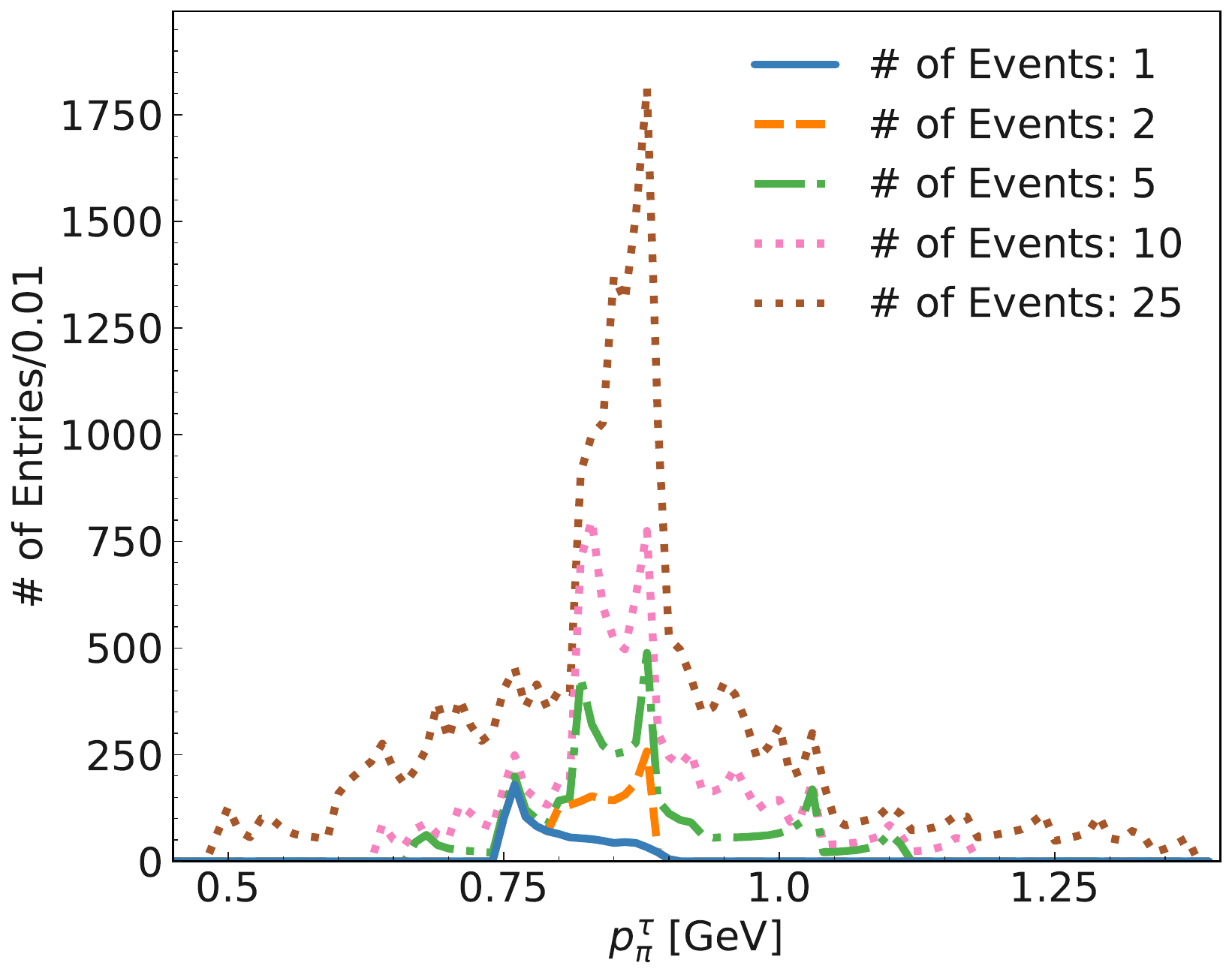}
    \caption{Step by step formation of the GKK limiting distribution for the $\tau \to \pi \nu$ decay. Here, for every \tautau event, we calculate a set of 1000 possible $\pi$-momentum candidates. The blue curve gives an example distribution for one event. The limiting distribution emerges in the plot by stacking many events, as indicated by the step-by-step distributions.}
    \label{fig:gkkFormation}
\end{figure}

In order to illustrate the capabilities of the GKK method in \tautau events, we consider the momentum spectrum of a $\tau^- \to \pi^- \nut$ decay, the signal-side, in the $\tau$-rest-frame. As described in Section~\ref{sec:concept}, we first sample a set of $p_{\tau}^{\mu}$ candidates of the tag-side $\tau$, considering the decay mode $\tau^- \to \pi^- \pi^+ \pi^- \nut$. 
We use the set of $p_{\tau}^{\mu}$ to boost the signal-side $\pi^-$-momentum, $p_{\pi}^{\mu}$, into the rest-frame of the signal $\tau$. We denote the momentum of the signal-side in the $\tau$-rest-frame as $p^{\tau}_{\pi}$.
In the process $\tau^- \to \pi^- \nut$, we expect a peak at $p^{\tau}_{\pi}=\frac{\m_{\tau} -m_{\pi}}{2}$. First, we illustrate in Figure~\ref{fig:gkkFormation} how the resulting GKK limiting distribution (in short, GKK-distribution) forms. We do this by stacking the $p_{\pi}^{\tau}$-distributions of each event.
As the number of events increases, a limiting distribution emerges, which should only depend on the parameter of interest (in our particular case, $p^{\tau}_{\pi}$). With 25 events, a clear peak emerges around the expected $p^{\tau}_{\pi}\approx 0.82\gev$.
If there are enough events and the input-mass for $m_{\tau}$ is the actual mass, as is the case with the blue line in Figure \ref{fig:tauSmear}, a sharp peak emerges at the expected momentum. We interpret this behaviour as a washed-out version of the actual distribution. Other examples for this may be found in~\cite{dissertation}.

%
\subsection{GKK: A New Method to Measure the $\tau$-Mass}
\label{ssec:2bodyDecay}
\begin{figure}[htbp]
	\subfloat[]{
		\begin{minipage}{0.49\linewidth}
			\includegraphics[width=0.95\linewidth]{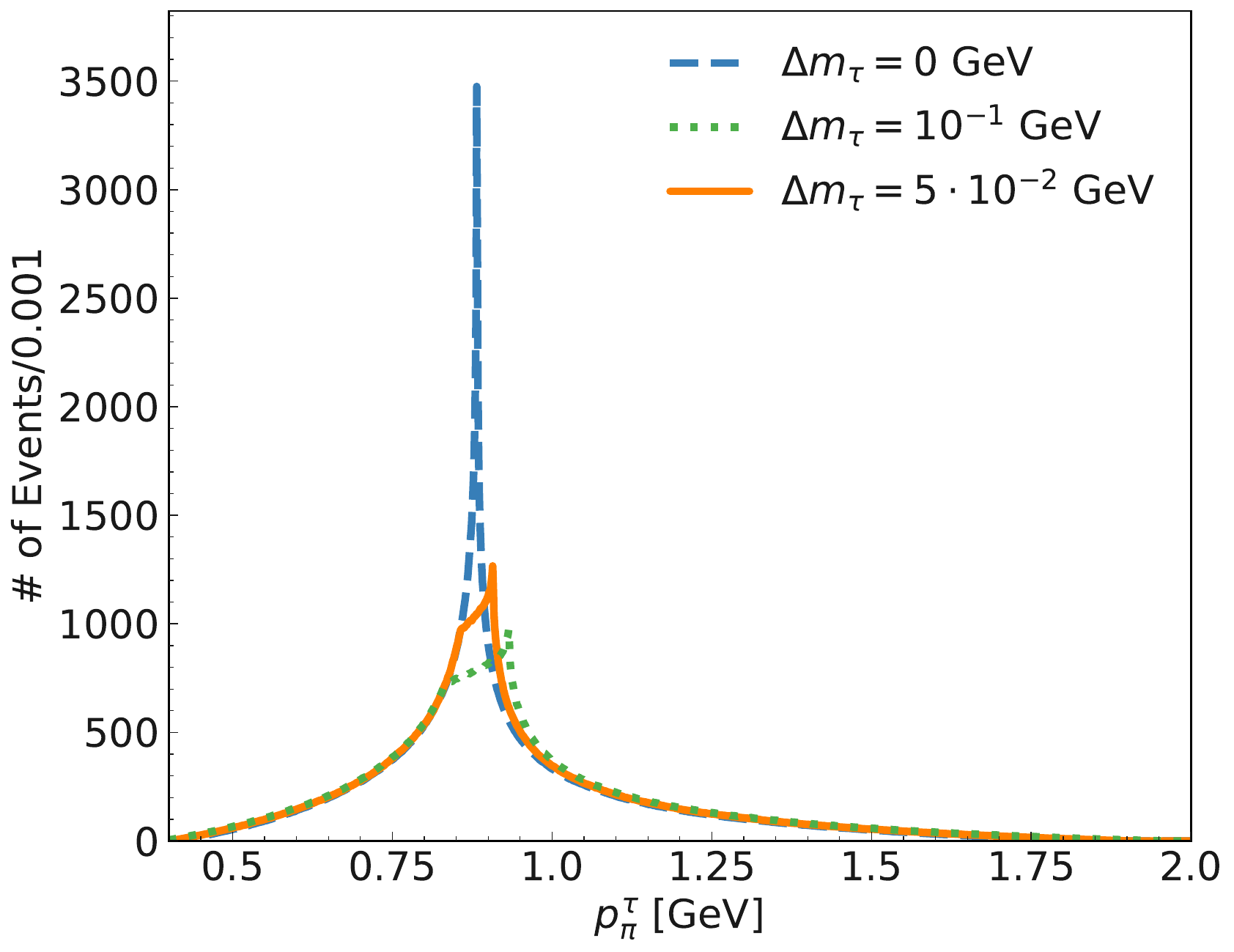}
			\label{fig:tauSmear}
		\end{minipage}
	}
	 \subfloat[]{
		\begin{minipage}{0.49\linewidth}
			\includegraphics[width=0.95\textwidth]{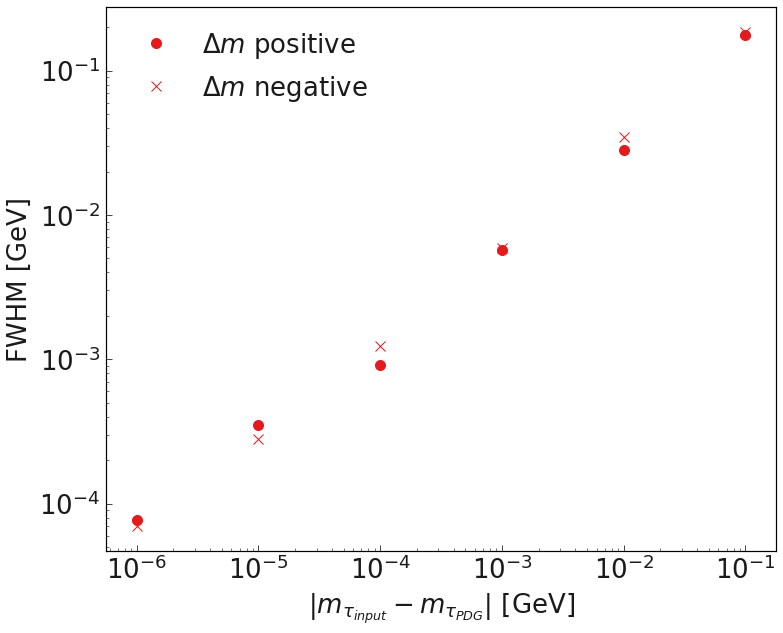}
			\label{fig:tauSmear2}
		\end{minipage}
	}
	\centering
	\caption{
The FWHM distribution illustrates the effect of the GKK distribution's smearing behaviour for different deviations of the mass hypothesis used to calculate the GKK distribution. 
	(a) GKK distribution for the $\pi$-momentum of the $\tau \to \pi \nu$ decay for $3\times10^5$ events in the $\tau$-rest-frame. 
	The effect of $\Delta m$ is demonstrated for the GKK distribution.
	(b) Full-Width Half Maximum (FWHM) for different $\Delta m$. 
	Here, Toy MC denotes simulation data.
	}
	\label{fig:tauToElSM}
\end{figure}

In order to use the GKK method for $\tau$-mass measurements, we need to understand the behaviour of the GKK method. We studied the influence of the $\tau$-mass input, which is needed to calculate the GKK distribution. This was done by considering the input masses $m_{\tau_\text{input}}$, which deviated by $\Delta m$ from the mass value  $m_{\tau_\text{PDG}}= 1776.86\pm0.12$~MeV~\cite{ParticleDataGroup:2020ssz} used in the event generation:
\begin{equation}
    \label{eq:deltaM}    
    \Delta m = m_{\tau_\text{input}} - m_{\tau_\text{PDG}}.
\end{equation} 
Figure~\ref{fig:tauSmear} illustrates the influence of $\Delta m$. Here, we compare the resulting limiting distribution for three different $\Delta m = {0, 0.05, 0.1}\gev$. To quantify the influence of $\Delta m$, we also calculate the Full-Width Half Maximum (FWHM) of the resulting limiting distributions. This quantity is a measure of the spread or smearing of the distribution. Figure \ref{fig:tauSmear2} shows the relation of FWHM versus $\Delta m$.

We observe two essential aspects of the GKK distribution from Figure~\ref{fig:tauSmear}. First, the true/expected momentum $p_{\pi, exp}^{\tau}$ is within the GKK-distributions peak region. We define the peak region between the two exponentially decreasing flanks of the distribution. 
Second, a simple peak search to determine the $\tau$-mass will not yield reliable results since a mismatch of the assumed and actual value of the mass, $\Delta m$, leads to a distortion of the peak shape. For non-zero $\Delta m$, the maximum of the distribution does not represent the actual $p_{\pi, exp}^{\tau}$. Instead, we observe that the width of the distribution increases with increasing $\Delta m$.

Figure~\ref{fig:tauSmear2} allows quantifying the dependence of the width on $\Delta m$, the Full-Width Half Maximum, FWHM, of the GKK-distribution for different values of $\Delta m$.
We calculate the FWHM of the distribution numerically. 
Here, the FWHM depends linearly on $\Delta m$ for negative and positive $\Delta m$ values. We do not observe any significant differences in the behaviour of a positive or negative $\Delta m$.

We interpret the GKK-distribution's behaviour as follows: the boost calculation incorporates the $\tau$-mass to compute the boost in the $\tau$-rest-frame and determine the candidate cone of the $\tau$-momenta. 
A mismatch between the true and the input value leads to a smearing effect in both cases. We use the distribution's width to quantify the smearing.

\begin{figure}
 \subfloat[]{
 \begin{minipage}{0.49\linewidth}
 \includegraphics[width=0.95\textwidth]{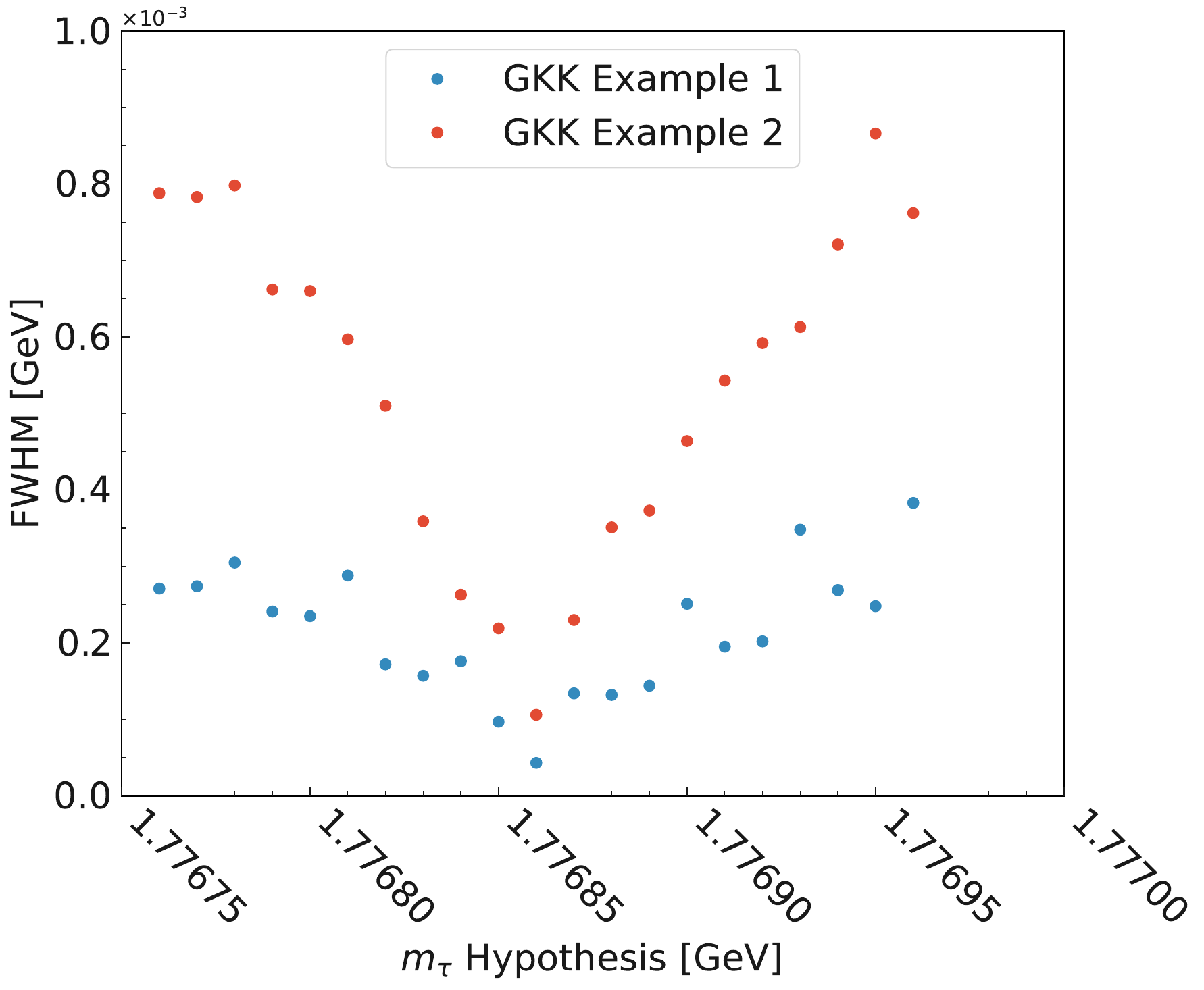}
 \label{fig:GKK_fwhm_example}
 \end{minipage}
 }
 \subfloat[]{
 \begin{minipage}{0.49\linewidth}
 \includegraphics[width=0.95\textwidth]{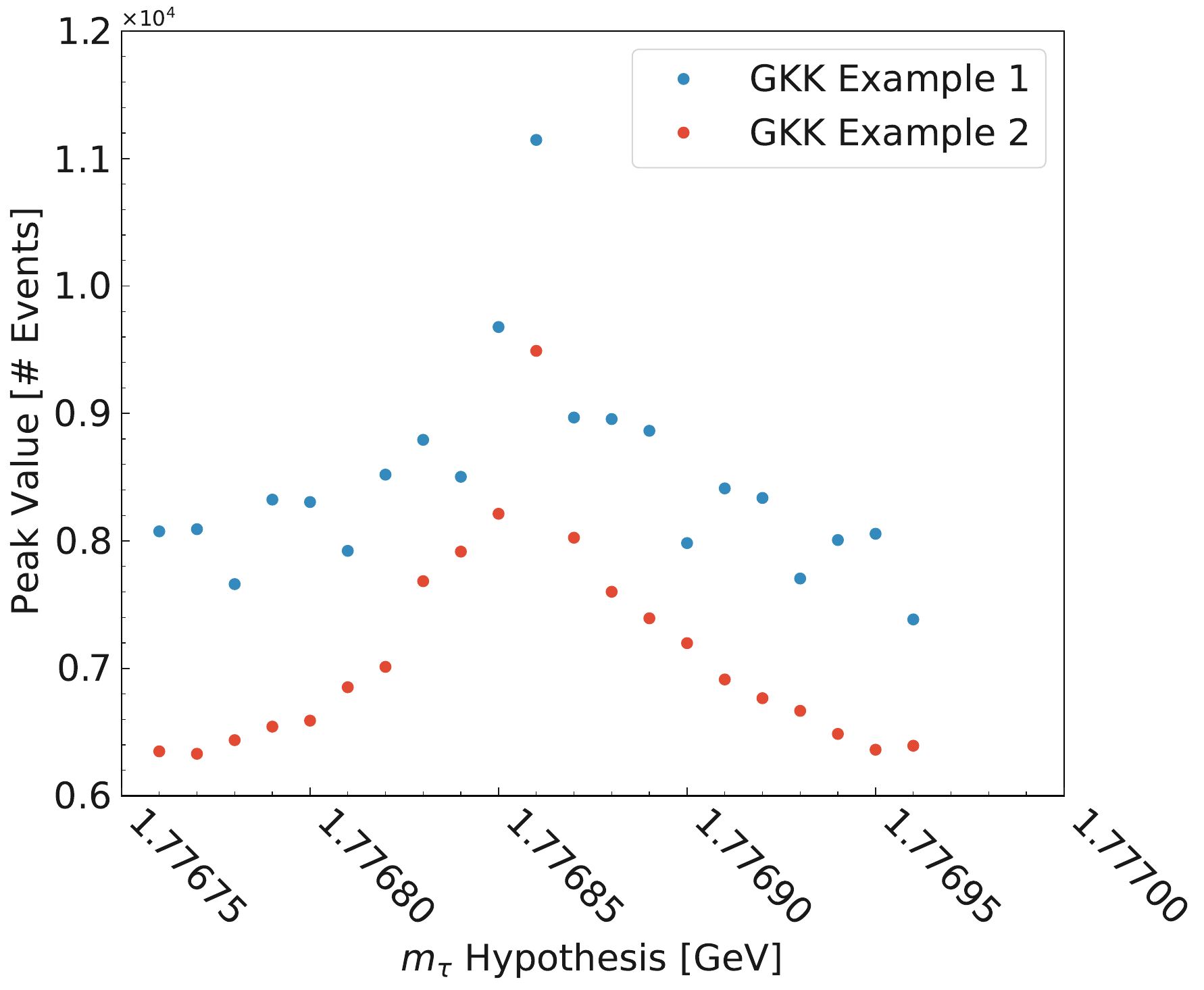}
 \label{fig:GKK_MaxPeak_example}
 \end{minipage}
 }\par
 \subfloat[]{
 \begin{minipage}{0.49\linewidth}
 \includegraphics[width=0.95\textwidth]{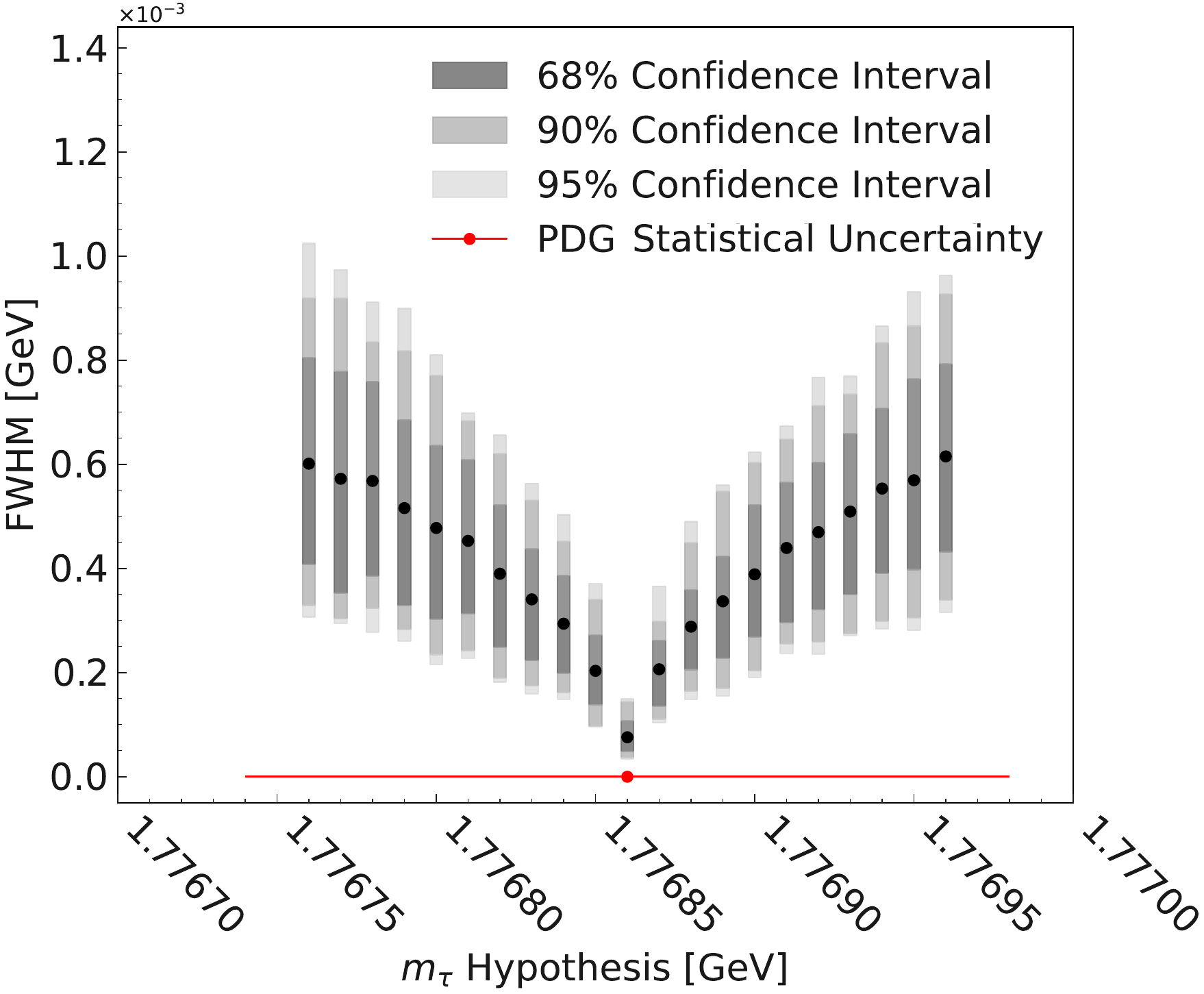}
 \label{fig:GKK_fwhm}
 \end{minipage}
}
\subfloat[]{
 \begin{minipage}{0.49\linewidth}
 \includegraphics[width=0.95\textwidth]{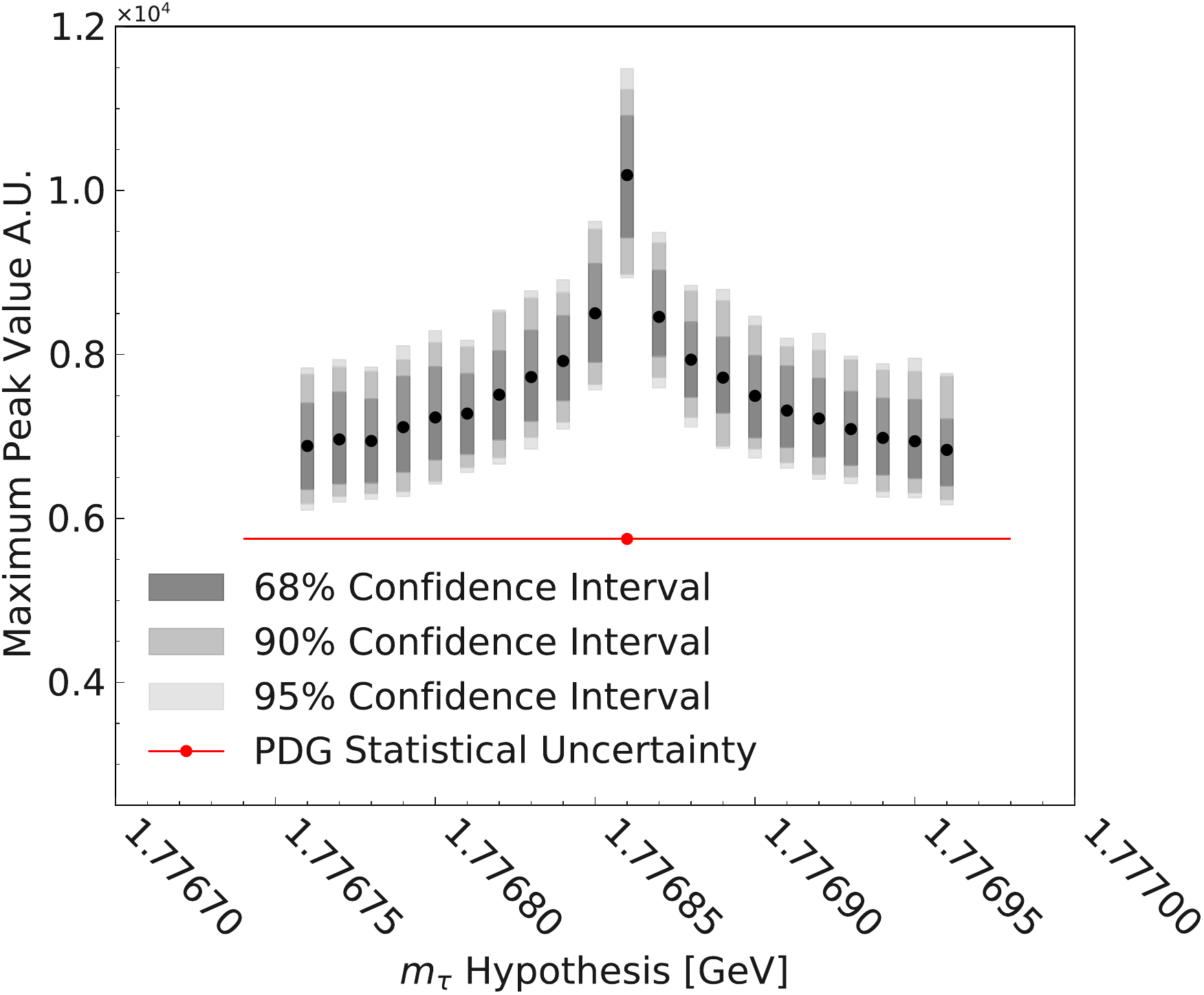}
 \label{fig:GKK_MaxPeak}
 \end{minipage}
}
 \centering
 \caption[FWHM and Max Peak Variation vs $m_{\tau}$ Calculation Inputs]{GKK-distribution's Full-Width Half Maximum (FWHM), plots (a) and (c), and Maximum Peak Value, plots (b) and (d), for a $\mtau$-mass hypothesis ($m_{\mtau_{\text{input}}}$) scanned around the initial simulation value of 1.77686\gevcc in 10\kevcc steps. 
 Plots (a) and (b) show the GKK distribution's FWHM and maximum peak value for two data sets with a shallow FWHM slope (Example 1) and a steep FWHM slope (Example 2), exemplifying the possible variations. The extrema for the FWHM and Maximum Peak Value are clearly at $m_{\mtau_{\text{PDG}}}$ for both data sets. Plots (c) and (d) show the confidence intervals indicating the expected slope variation for this scan's FWHM or Maximum Peak value. The PDG Statistical Uncertainty indicates the world average (dot) and the corresponding lowest statistical uncertainty measured by BaBar~\cite{BaBar:2009qmj} (red line). The displayed uncertainty also coincides with the total uncertainty of the PDG's average mass value.}
 \label{fig:GKK_linTauMass}
 \end{figure}

This result means it is possible to determine the $\tau$-mass by scanning through $m_{\tau_\text{input}}$ hypotheses. 
For example, we can determine the $\tau$-mass by minimising the FWHM in a numerical approach with different $m_{\tau_\text{input}}$. 
Figure~\ref{fig:GKK_fwhm_example} illustrates two examples of the FWHM distribution close to the actual simulation value. 
The distribution minimum corresponds to $m_{\tau_\text{PDG}}$.
Furthermore, we have found that the maximum peak value of the GKK distributions can also be used to extract the $\tau$-mass. Again, Figure~\ref{fig:GKK_MaxPeak_example} clearly shows that the extremum is at $m_{\tau_\text{PDG}}$ for both examples. Figures~\ref{fig:GKK_fwhm} and \ref{fig:GKK_MaxPeak} show the mean, as black dots, and the 68\%, 90\% and 95\% confidence levels, as bands, produced from 100 independent toy simulations without detector resolution\footnote{Please note that the $m_{\tau_\text{input}}$ scan yields a distribution as indicated by the examples in figures~\ref{fig:GKK_fwhm_example} and \ref{fig:GKK_MaxPeak_example} or the black dots. The confidence level shows the variance in the width of the FWHM or maximum peak value distribution.}. 
We note that the GKK method clearly distinguishes the $m_{\mtau_{\text{PDG}}}$ in the plots, indicating a precision of at least the step size of 10\kev for this method. All 100 toy simulations showed an extremum in the same peak, indicating the result's high robustness. The red dot with the error bar indicates the PDG average and corresponding (statistical) uncertainty for the $m_{\tau}$ mass. With the generated data set, the GKK method is less limited by statistical uncertainty than methods used in the past. We expect that systematic uncertainties will dominate future measurements.

\subsection{Sensitivity Study with Realistic Detector Smearing} 
\begin{figure}[htbp]
	\subfloat[]{
		\begin{minipage}{0.49\linewidth}
			\includegraphics[width=0.95\linewidth]{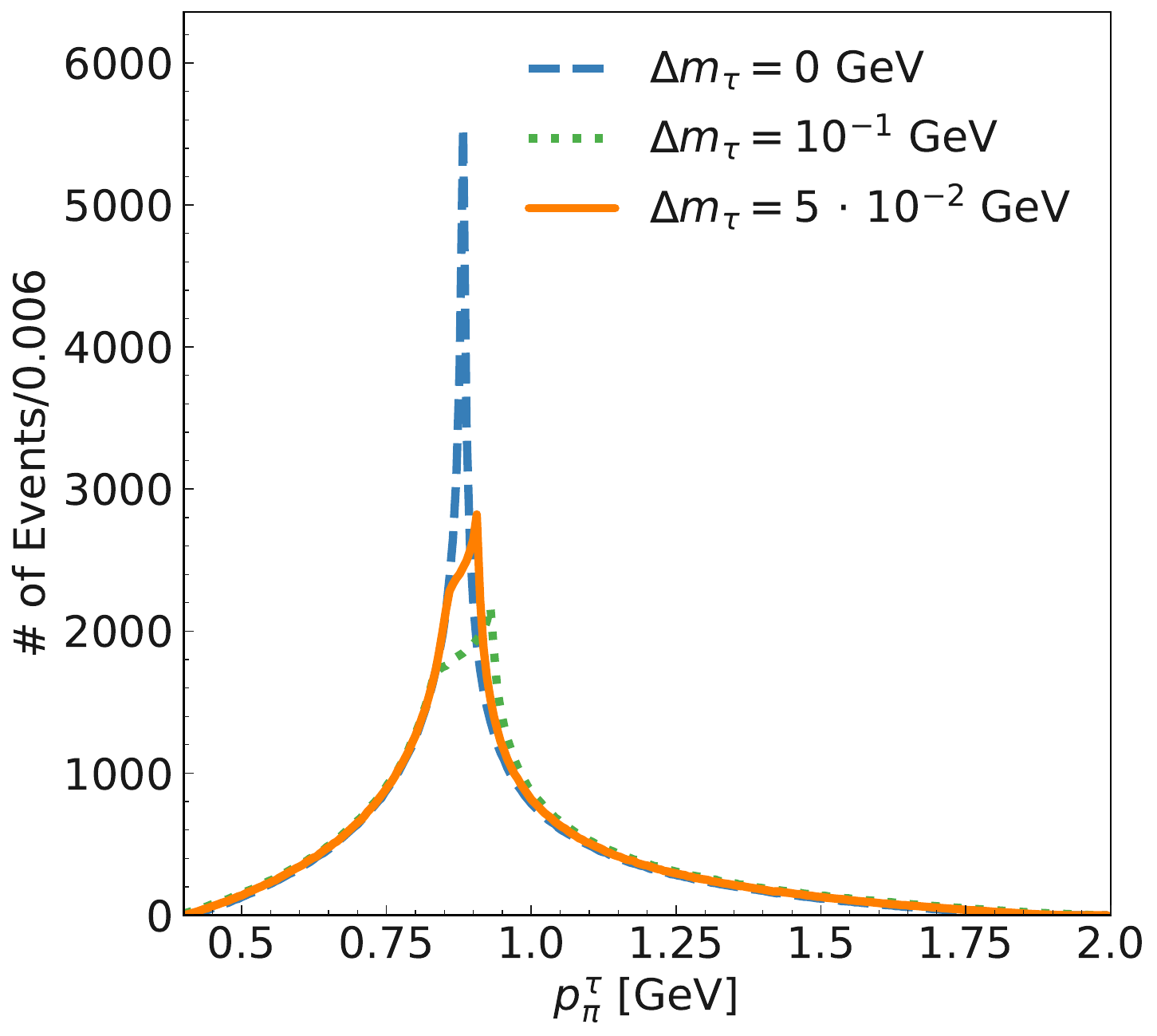}
			\label{fig:tauSmear_detector}
		\end{minipage}
	}
	 \subfloat[]{
		\begin{minipage}{0.49\linewidth}
			\includegraphics[width=0.95\textwidth]{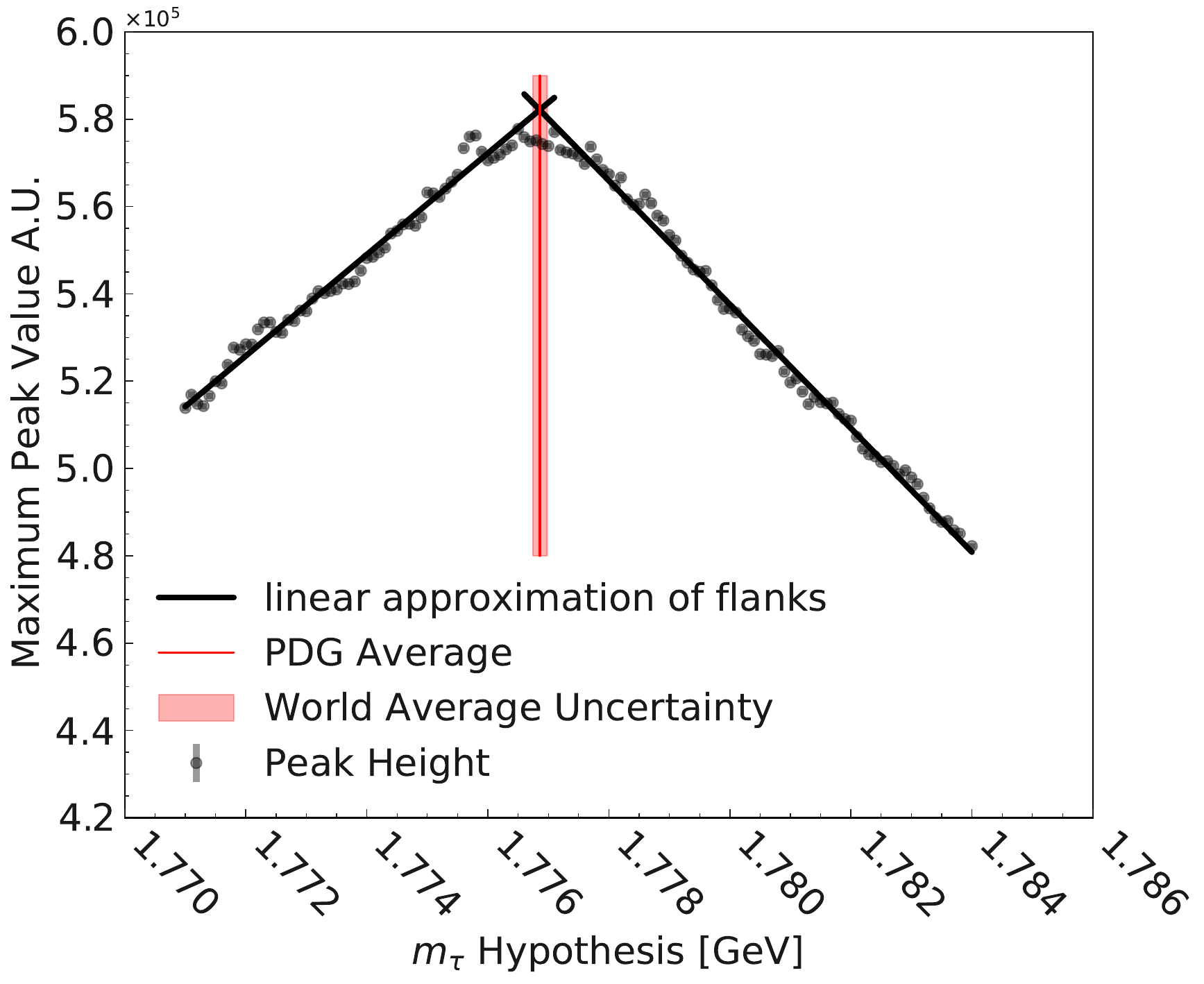}
			\label{fig:tauSmear_detector2}
		\end{minipage}
	}
	\centering
	\caption{ The Maximum Value distribution illustrates the effect of the GKK-distribution's smearing behaviour for different deviations of the mass hypothesis used to calculate the GKK distribution. 
	(a) GKK-distribution for the $\pi$-momentum of the $\tau \to \pi \nu$ decay for $1.2\times10^6$ events in the $\tau$-rest-frame with simulated detector effects. 
	The effect of $\Delta m$ is demonstrated for the GKK distribution.
	(b) Maximum Value for different $m_{\tau}$ hypotheses with statistical uncertainty, see text for details. We fit the decreasing left and right parts of the Maximum Value distribution with two linear functions (black lines). Our estimate for $m_{\tau}$ is the intersection of the two linear functions; see text for details.
	}
	\label{fig:tauToElSM_detector}
\end{figure}

To assess the capability of estimating the $\tau$-mass, we simulated detector effects by introducing Gaussian smearing according to the momentum resolution of state-of-the-art detector systems~\cite{BelleIITrackingGroup:2020hpx}. We consider a data set of 1.2 million signal $\tautau$ events, corresponding to an integrated luminosity of about 500~fb$^-1$ at the $\Upsilon(4S)$ resonance. We estimate the reconstruction efficiency with about 15\% according to a recently published result~\cite{Belle-II:2022heu}. Figure~\ref{fig:tauSmear_detector} displays the resulting GKK-distribution when considering detector effects in equivalence to Figure~\ref{fig:tauSmear}.

We observe that the detector resolution effects cause a broadening of the GKK distribution when comparing the blue dashed lines of figures~\ref{fig:tauSmear} and \ref{fig:tauSmear_detector}. When comparing the blue-dashed lines with the orange and green dotted lines in Figure~\ref{fig:tauSmear_detector}, we observe that the effect of $\Delta m$ is retained for the GKK distribution. 

We estimate the sensitivity of the $m_{\tau}$ measurement using the Maximum Value, $n^{\text{max}}_b$ distribution. In this case, the uncertainty of the numerically determined value is given by the Poisson error of the candidates in the bin.
We model the increasing and decreasing parts of the Maximum Value distribution with two linear functions, respectively, because this is a plausible first-order approximation. We estimate $m_{\tau}$ as the peak of the Maximum Value distribution and determine this peak as the intersection of the two linear functions. Our estimate is:
\begin{equation}
    m_{\tau} = 1776.86 \pm 0.09~\text{[MeV]},
\end{equation}
in agreement with the input value of $1776.86~\text{[MeV]}$

Currently, the ARGUS method is the leading method for measuring the $\tau$-mass at \epem-colliders well above the \tautau-production threshold \cite{Belle:2006qqw, BaBar:2009qmj}. It has worse precision than the one used by the BES III collaboration exploring the \tautau-production threshold~\cite{Anashin:2007zz, BESIII:2014srs}.

The method proposed here uses the information of all available events in the chosen decay topology for measuring the $\tau$-mass. This property is an advantage compared to the ARGUS method, which uses only the subsample of the events close to its distributions endpoint~\cite{ARGUS:1992chv}. 
Moreover, the ARGUS method has an intrinsic bias, which must be corrected based on simulation studies. In contrast, the GKK method proposed here provides a direct estimate for the $\tau$-mass without the need for further method-based corrections.
We see the potential that the GKK method can improve the accuracy and precision of the $\tau$-mass measurement compared to the ARGUS method. At the very least, it provides a complementary approach to the current method.

\section{Conclusions}
We have presented a new method for determining observables in particle-pair events. Using an explicit example, we demonstrated the method as \tautau events in \epem collisions. We showed that inferring physical constraints for missing information leads to a probability distribution that can be treated numerically and shows the properties of an unbiased best estimator. We demonstrated the example of the $\tau$-mass measurement as a concrete use case of the GKK-method, where the mass can be extracted without the need of further method based corrections. Moreover, this method can determine the mass of a new physics particle in $\tau$-events. 
Here, we turn around the argument. Instead of two fixed daughter-particle masses, we replace one with the known mother-particle mass. The peak position determines the unknown daughter-particle mass, making the GKK method suitable for a massive invisible particle search.

We showed that the GKK-method could lead to an improvement in parameter estimation. The $\tau$-mass example displayed an apparent linear behaviour enabling a future precise measurement. Further studies which evaluate the performance in an actual detector environment are necessary to evaluate the possible improvements over present-day techniques.

Furthermore, other properties of the GKK method remain to be studied. We expect an analytic description of the GKK distribution's limiting distribution. A possible function family could be the Asymmetric Generalised Gaussian Family of Distributions \cite{778737}. With an analytic description, we believe that the parameter estimation and smearing could be described more reliably than with the numeric approach used until now. 

%
\bibliographystyle{unsrt}
\bibliography{bibfile}
%
\end{document}